\documentclass{aa}                        
\usepackage{graphicx, latexsym}                   
\usepackage{natbib}                       
\bibpunct{(}{)}{;}{a}{}{,}               

\defcitealias{geo05}{Paper~I}

\begin{document}

\title{2D non-LTE Modeling for Axi-symmetric Winds. II. A Short
Characteristic Solution for Radiative Transfer in Rotating Winds}

\author{J. Zsarg\'{o}\inst{1}
\and D. J. Hillier\inst{1}
\and L. N. Georgiev\inst{2}\fnmsep\inst{1}}

\institute{ Dept. of Physics and Astronomy,
University of Pittsburgh, 3941 O'Hara St.,
Pittsburgh, PA 15260, USA
\and \emph{Present Address:} Instituto de Astronomia,
Universidad Nacional Autonoma de Mexico (UNAM),
CD. Universitaria, Apartado Postal 70-264,
04510, M\'{e}xico DF, M\'{e}xico}

\offprints{J.Zsarg\'{o}, \email{jzsargo@astro.phyast.pitt.edu}}

\date{Received 30 June 2005/ Accepted 4 October 2005}

\abstract{
We present a new radiative transfer code for axi-symmetric
stellar atmospheres and compare test results against 1D and 2D
models with and without velocity fields.
The code uses the short characteristic method with modifications to
handle axi-symmetric and non-monotonic 3D wind velocities, and allows
for distributed calculations.
The formal solution along a characteristic is evaluated with a resolution
that is proportional to the velocity gradient along the characteristic.
This allows us to accurately map the variation of the opacities and
emissivities as a function of frequency and spatial coordinates, but
avoids unnecessary work in low velocity regions.
We represent a characteristic with an impact-parameter vector {\bf p}
(a vector that is normal to the plane containing the characteristic and the origin)
rather than the traditional unit vector in the direction of the ray.
The code calculates the incoming intensities for the characteristics by a
single latitudinal interpolation without any further interpolation in the
radiation angles.
Using  this representation also provides a venue for distributed
calculations since the radiative transfer can be done independently for each
{\bf p}.
}

\titlerunning{A SC Solution for Radiative Transfer in Rotating Winds}
\authorrunning{Zsarg\'{o} et al.}

\maketitle

\section{Introduction}\label{section:intro}

Massive stars and their winds play an important role in shaping the dynamical
structure and energy budget of galaxies.
For example, they enrich the ISM with nuclear processed material and deposit
large amounts of mechanical energy into their surroundings.
Despite decades of research and considerable advancements in our understanding
of stellar envelopes, there is still much to learn.
Because of the complexities of these systems, and the increasing emphasis on the
details, it has become very difficult to proceed without complex numerical simulations.
It is not surprising, therefore, that the history of stellar studies reflects not only our
advancing knowledge but also our increasing computational capabilities.
Initially, simple plane-parallel LTE models were utilized in numerical
simulations \citep[see e.g.,][and references therein]{kur91} and these were
adequate for stars with dense atmospheres and low mass-loss rates.
These models were also the only simulations that were viable on the computing
facilities of the time.
Unfortunately, the above simplifications cannot be extended to most early-type stars.
\cite{aue72, aue73}, for example, demonstrated that the assumption of LTE
is invalid in O-type stars and the statistical equilibrium equations need to be
solved for the level populations.
For massive stars with extensive mass-loss (e.g., Wolf-Rayet stars) geometrical
effects are also important and plane-parallel models are no longer sufficient.
As a minimum, therefore, one needs to use non-LTE spherical models to understand
these objects.
The system of statistical equilibrium equations, however, is highly non-linear in the level
populations and finding a solution for fully line blanketed models
is a formidable task.
We have reached the necessary level in computing power only in the last few years
to be able to routinely perform such computations
\citep[see e.g.,][]{hub95, hau96, hil98, pau01, gra02}.

Plane-parallel and spherical non-LTE modeling have found wide applicability
in spectroscopic studies.
Recent works by \cite{mar02, cro02, hil03, her02} have revised the temperature
scale for O stars, for example, and have given new insights into the structure of stellar
winds.
However, spherical (or plane-parallel) modeling also has its limitations and
cannot be used to study many important stellar objects.

It has been known for a long time that some circumstellar envelopes are non-spherical
--- the most well-known examples are the envelopes of Be stars.
The hydrogen emission and infrared excess of these stars are thought to be produced in a
thin disk.
The presence of these disks was inferred from both line modeling and
polarimetric studies \citep{poe78a, poe78b}, and has been confirmed by interferometric
observations \citep{ste95, qui97}.
Furthermore, recent MHD simulations \citep{cas02, udd02, owo04} argue for
equatorial confinement by magnetic field for the origin of the disks.
If a dynamically important magnetic field is present in Be envelopes that in itself
ensures at least a 2D nature of their wind.

Other stellar problems for which 1D models are inadequate include rapidly
rotating OB stars, binaries with colliding winds or accretion disks,
pre-main sequence and young stars, stellar envelopes irradiated by external sources
(e.g., massive stars near an AGN), and the collapsing core (Type-II) supernovae
\citep[e.g.,][]{wan01, kif03}.
Advanced supernovae models may even have cosmological applications since
these luminous objects can be used as distance calibrators in the nearby
universe \citep[see][and references therein]{des05a, des05b}.

The case of rapid OB rotators is particularly important for this paper
since we test our code on such a problem.
These stars are subjects of intense research and the exact structure of the
rotating envelope is not well established.
The conservation of angular momentum in the wind may result in
meridional flow toward the equator which potentially leads to disk formation
\citep[see e.g.,][]{bjo93}.  Conversely the latitudinal
variation of the surface gravity will result in a variation of the radiative flux
with latitude that can inhibit disk formation, and can cause
a strong polar wind \citep{owo96, mae00}.
Either way, the underlying spherical symmetry of the outflow is broken and
at least axi-symmetric models are needed for spectral analysis.

Motivated by the need for 2D model atmospheres, and by the availability
of fast computers and methods, we undertook a project to develop
a tool for spectroscopic analysis of axi-symmetric stellar envelopes.
The solution of the statistical equilibrium equations for the level
populations and temperature is discussed in the first paper of
this series \citep[][Paper~I]{geo05}.
At present the main code, ASTAROTH, solves for the radiation field by a continuum
transfer routine that is based on the method of \cite{bus00} and uses the Sobolev
approximation for line transfer.
In this paper we present an alternate routine for ASTAROTH that can handle the
line-transfer without the use of Sobolev approximation in models
with continuous, but not necessarily monotonic, velocity fields.
We treated this problem independently from the main project because
it required experimentation with alternate solution methods.
In \S\ref{section:code} we describe our goals and motivations
in finding the proper solution method, and we also give a brief discussion
of the chosen approach.
The C++ code that was developed for
the transfer is described in \S\ref{section:tests} where we also present
the test results and verification.
Finally, we draw our conclusions in \S\ref{section:con}.

\section{Description of the Solution Technique}
\label{section:code}

A non-LTE model of a stellar envelope is a complex nonlinear problem.
The level populations and the radiation field are strongly coupled.
Thus, an iterative procedure is needed to achieve a consistent solution.
To solve the statistical equilibrium equations for the level populations, one
must determine the radiative transition rates for free-free, bound-free and
bound-bound transitions.
These require the knowledge of the radiation moments
\begin{equation}\label{eq:J}
J({\bf r}, \nu) = \frac{1}{4 \pi} \int_{\Omega} I({\bf r}, \underline{\bf n}, \nu)
\; d\Omega
\end{equation}
and
\begin{equation}\label{eq:Jbar}
\overline{J}_l ({\bf r}) = \frac{1}{4 \pi} \int_{\Omega} \int_{0}^{\infty}
 I({\bf r}, \underline{\bf n}, \nu) \Phi_l (\nu) \; d\nu  d\Omega \;\;  .
\end{equation}
The quantities $I({\bf r}, \underline{\bf n}, \nu) $, {\bf r}, and $\underline{\bf n}$ are
the specific intensity, the spatial position, and the direction in which the radiation is
propagating, respectively.
The function $\Phi_l$ represents the normalized line-profile for any given bound-bound
transition and the integrations are over all solid angles and frequencies.

Only $J$ and $\overline{J}_l$ are needed to solve the statistical
equilibrium equations, but they have to be updated every iteration cycle.
This introduces stringent requirements on numerical efficiency and speed, but also
allows for simplifications.
The Radiative Transfer (RT) code does not have to produce the observed spectrum,
for example, since it is irrelevant for the transition rates. Nor do
the specific intensities at each depth need to be stored.
On the other hand, the run time characteristics of the code are critical for its
application in an iterative procedure.
Therefore, our RT code is optimized to calculate $J$, $\overline{J}_l$,
and the ``approximate lambda operator'' ($\Lambda^*$, see \S\ref{section:ALO})
as efficiently as possible.
Crude spectra in the observer's frame are calculated only if requested,
and only for monitoring the behavior of the code.

At a minimum, a realistic non-LTE and line-blanketed model
atmosphere requires the inclusion of most H, He, C, N, O, and
a large fraction of Fe transitions in the calculation.
The running time and memory requirements of such a model
can be several orders of magnitude larger in 2D than those of its spherical or
plane-parallel counterpart.
The dramatic increase in computational
effort arises from both the extra spatial dimension, and from the
extra variable needed to describe the angular variation of the
radiation field.
In spherical models, for example, the radiation field is symmetric
around the radial direction --- a symmetry which is lost in 2D.
We believe that realistic 2D/3D simulations, especially in the presence of
non-monotonic flow velocities, will inevitably require the simultaneous use
of multiple processors.
Therefore, we developed ASTAROTH and this RT code to be suitable
for distributed calculations by ensuring that their sub-tasks are as independent
from each other as possible.

\subsection{The Solution of the Radiative Transfer}
\label{section:solution}

Our choice to calculate moments
$J$ and $\overline{J}_l$ is to solve the radiative
transfer equation for static and non-relativistic media
\begin{equation}\label{eq:RT}
 \underline{\bf n}  {\bf \nabla}  I({\bf r}, \underline{\bf n}, \nu)=
 - \chi({\bf r}, \underline{\bf n}, \nu) \left[ I({\bf r}, \underline{\bf n}, \nu) -
 S({\bf r}, \underline{\bf n}, \nu) \right] \; ,
\end{equation}
and then evaluate the integrals in Eqs.~\ref{eq:J} and \ref{eq:Jbar}.
The quantities $\chi$ and $S$ in Eq.~\ref{eq:RT} are the opacity and
source function, respectively.
A major simplification in this approach is that a formal solution
exists for Eq.~\ref{eq:RT}.
At any $s$ position along a given ray (or characteristic), the optical depth and
the specific intensity are
\begin{equation}\label{eq:tau}
\tau_{\nu}= \int_{0}^{s} \chi ds'
\end{equation}
and
\begin{equation}\label{eq:I}
I(\tau_{\nu})= I_{BC} \, e^{- \tau_{\nu}} \; + \;  \int_{0}^{\tau_{\nu}} S(\tau') \,
e^{\tau' - \tau_{\nu}} \, d\tau'  \; ,
\end{equation}
respectively (from now on, we stop indicating functional dependence of quantities on
{\bf r}, $\underline{\bf n}$, and $\nu$).
Therefore, the intensity can be calculated by specifying $I_{BC}$ at the
up-stream end ($s=0$) of the ray  and by evaluating two integrals
(assuming that $S$ and $\chi$ are known).
We sample the radiation field by a number of rays for every
spatial point.
If the number and the orientation of the rays are chosen properly, then the
angular variation of $I$ is sufficiently reproduced and accurate
$J$ and $\overline{J}_l$ can be calculated.

There are alternatives to this simple approach; each has its own merits and
drawbacks.
For example, from Eq.~\ref{eq:RT} one can derive differential equations 
for the moments of the radiation field and solve for them directly.
This approach has been successfully used in 1D codes, like CMFGEN \citep{hil98},
and in calculations for 2D continuum/grey problems \citep{bus01}.
A distinct advantage of the method is that electron scattering (ES) is
explicitly included in the equations, and consequently no ES iteration is
needed.
However, to achieve a closed system of moment equations a closure relationship
between the various moments is required.
This relationship is generally derived from the formal solution which requires at
least a fast and rudimentary evaluation of Eqs.~\ref{eq:tau} and \ref{eq:I}.
Furthermore, the 2D moment equations are quite complicated and it is not easy
to formulate the proper boundary conditions in the presence of non-monotonic
velocity fields.
For our purposes we needed a simple approach that is flexible enough to
implement in distributed calculations.

An increasingly popular method to solve the RT is using Monte-Carlo
simulations. In this method, a large number of photon packets
are followed through the envelope and the properties of the radiation field
are estimated by using this photon ensemble \citep[see e.g.,][]{luc99, luc02, luc03}.
While the Monte-Carlo simulations are flexible and suitable for parallel computing,
they can also have undesirable run-time characteristics.
It is also unclear how line overlaps in the presence of a non-monotonic
velocity field can be treated by Monte-Carlo techniques without the use of Sobolev
approximation.

After considering our needs and options, we decided to use the straightforward
approach, solving Eq.~\ref{eq:RT} and evaluating Eqs.~\ref{eq:J} and \ref{eq:Jbar}.
This approach provides a reasonable compromise of accuracy, numerical efficiency,
and flexibility.
Our code will also increase the pool of available RT programs in stellar studies.
Each solution technique has its specific strength (e.g., our method is fast enough
for an iterative procedure) and weaknesses; therefore, future researchers will have
more options to choose the best method for their needs.
Having a selection of RT codes that are based on different solution methods will
also allow for appropriate cross-checking of newly developed programs.
\begin{figure}
\resizebox{\hsize}{!}{
\includegraphics[angle= 270]{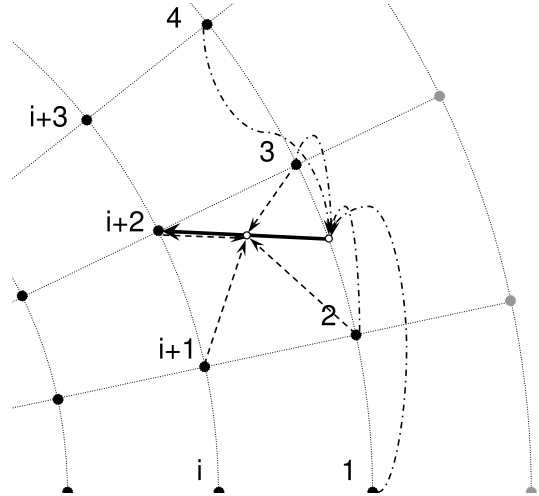}
}
\caption{
A sub-section of a typical spatial grid used in our RT code.
The boundary and internal points are indicated by grey
and black dots, respectively.
The solid arrow represents a SC belonging to point i+2
and pointing in the direction of the radiation.
Note, that the characteristic is terminated at the closest cell
boundary (between nodes 2 and 3), and is not followed all the way
to the boundary of the domain (grey points).
The numbering at the nodes indicates the order in which
the intensity in this direction is evaluated.
The small empty circles on the SC are the integration points (see
\S\ref{section:solution}) and the dashed arrows show which grid
points are used for interpolating $\chi$ and $S$ (straight arrows),
or $I_{BC}$ (curved arrows).
}
\label{fig1}
\end{figure}

The most accurate solutions for Eqs.~\ref{eq:tau} and \ref{eq:I} are achieved
when the integrals are evaluated all the way to the boundary of the modeling domain
along each ray \citep[Long Characteristic (LC) method,][]{jon73a, jon73b}.
To increase efficiency, we decided to use the so-called ``Short-Characteristic''
(SC) method, first explored by \cite{mih78} and \cite{kun88}.
In our implementation of this method, the characteristics are terminated at
the next up-stream radial shell (normally, they would be terminated at any
cell boundary) where $I_{BC}$  is calculated by an interpolation
between the specific intensities of the nearest latitudinal grid points
(see Fig.~\ref{fig1}).
We calculate the specific intensity in a given direction for all grid points starting
with those at the upstream end of the domain (where $I$ is set to the appropriate
boundary condition) and proceed with the calculation downstream (see Fig.~\ref{fig1}
for details).
This evaluation scheme ensures that all intensity values are calculated
by the time they are needed for the interpolation of $I_{BC}$.
With this simple trick, the specific intensity is calculated very efficiently
but for the cost of introducing coupling between the directional
sampling of the intensity at the grid points.
We will discuss the implications of this coupling in \S\ref{section:dir}.

On every SC, we evaluate the integrals of Eqs.~\ref{eq:tau} and \ref{eq:I} for every
co-moving frequency of the down-stream end point  ($i$+2 in Fig~\ref{fig1}) by
\begin{eqnarray}\label{eq:inttau}
\tau= \sum_{j=1}^{N-1} \Delta \tau_j   & ~~~~~ &  \Delta \tau_j=  \frac{\chi_{j+1} +
\chi_j}{2} (s_{j+1} - s_j)
\end{eqnarray}
and
\begin{eqnarray}
 \int_{0}^{\tau_{\nu}} S(\tau') \, && e^{\tau' - \tau_{\nu}} \, d\tau'  = 
 \sum_{j=1}^{N-1}
 \frac{S_{j+1}}{\Delta \tau_j} \left( \Delta \tau_j + e^{- \Delta \tau_j }
  - 1 \right) \nonumber  \\
 & & - \sum_{j=1}^{N-1}  \frac{S_{j}}{\Delta \tau_j}  \left( \Delta \tau_j + \left( 1 +
   \Delta \tau_j \right) \left( e^{- \Delta \tau_j} - 1 \right) \right)
\label{eq:intS}
\end{eqnarray}
where $N$-1 is the number of integration steps.
Eqs.~\ref{eq:inttau} and \ref{eq:intS} can be easily derived from Eqs.~\ref{eq:tau}
and \ref{eq:I} by assuming that in each interval $\chi$ and $S$ are linear in $s$ and
$\tau$, respectively.
To ensure that the spatial and frequency variations of the opacity and source
function are mapped properly, we divide the SC into small $s_{j+1} - s_j$ intervals
by placing enough ``integration'' points on the characteristic.
The number of these points ($N$) depends on the ratio of the ``maximum line
of sight velocity difference'' along the SC and an adjustable ``maximum allowed
velocity difference''.
By choosing this free parameter properly we ensure adequate frequency mapping
but avoid unnecessary calculations in low velocity regions.
Further, we can trade accuracy for speed at the early stages of the iteration and later
``slow down'' for accuracy.
We allowed for 20~km~s$^{-1}$ velocity differences along any SC in the calculations
that we present here.
Even though this is larger than the average frequency resolution of our opacity
and emissivity data ($\sim$10~km~s$^{-1}$), it was still adequate.
Trial runs with  2~km~s$^{-1}$ and 20~km~s$^{-1}$ ``maximum allowed velocity difference''
for the 1D model with realistic wind velocities (see \S\ref{section:1Dwind}) produced
nearly identical results.

The line of sight velocities, $\chi_j$, and $S_j$  are calculated at
the integration points by bi-linear interpolations using the four closest spatial grid
points (see Appendix~\ref{section:appendixA} and Fig.~\ref{fig1}).
We would like to emphasize, that the interpolated $\chi_j$ and $S_j$ are in the
co-moving frame and not in the frame in which the integration is performed.
This difference must be taken into account in Eqs.~\ref{eq:inttau}--\ref{eq:intS}
by applying the proper Doppler shifts at each integration point
(see Appendix~\ref{section:appendixA}).

With the exception of the intensity, all quantities are interpolated assuming
that they vary linearly between nodes.
Extensive testing of our code revealed that at least a third-order interpolation
is necessary to calculate $I_{BC}$ sufficiently accurately
(see Appendix~\ref{section:appendixB}).
For all other quantities first-order approximation is adequate in most cases but
not in all.
Since we wished to keep the first-order approximations if possible
(it is the least time consuming and is numerically well behaved),
a simple multi-grid approach was introduced to improve accuracy.
Unlike the intensity calculation, the interpolation of $\chi$ and $S$ 
{\em do not} have to be performed on the main grid;
therefore, a dense spatial grid for opacities and source functions
can be created, using monotonic cubic interpolation \citep{ste90},
before the start of the calculation.
Then, we use this dense grid to perform the bi-linear interpolations
to the integration points but perform the RT calculation only for
spatial points on the main grid.
Before the next iteration, the opacities and source terms on the dense grid
are updated.
To ensure a straightforward $\Lambda^*$ calculation we require the main grid
to be a sub-grid of the dense grid.
Further, the use of the dense grid is optional and only required if more
accurate approximations of $\chi$ and $S$ are desired.
With this rudimentary multi-grid technique, we improved the accuracy of
our calculations for essentially no cost in running time ($\sim$5-10\% increase).
However, there was a substantial increase in memory requirement.
To avoid depleting the available memory, the RT is usually
performed in frequency batches that can be tailored to fit into the
available memory.
This technique not only decreases the memory requirements, but also
provides an excellent opportunity for parallelization.

\subsection{Our Coordinate System and Representation of
Directions}\label{section:dir}

\begin{figure}
\resizebox{\hsize}{!}{
\includegraphics[angle= 270]{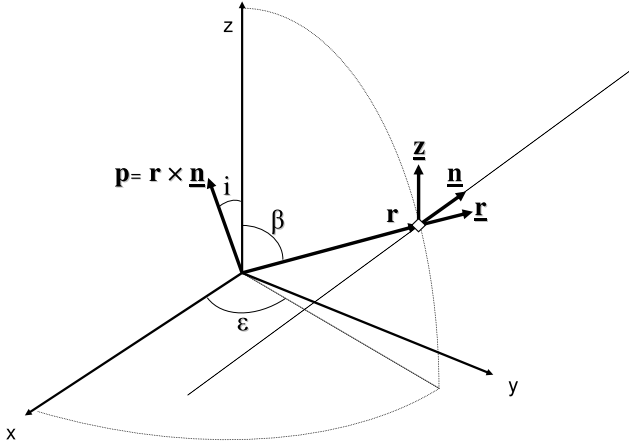}
}
\caption{
The definition of our fundamental coordinate system.
The unit vector $\underline{\bf n}$ describes a characteristic (long
thin line) pointing in the direction of the radiation and $r$, $\beta$, and
$\epsilon$ are the traditional polar coordinates of a spatial point.
Note that it is assumed here and in the rest of the paper that $z$ axis is
the axis of symmetry.
We use the impact-parameter vector {\bf p} (which is perpendicular to the
plane containing the characteristic and the origin), instead of
$\underline{\bf n}$, to represent a particular characteristic
(see \S\ref{section:dir} for explanations).
}
\label{fig2}
\end{figure}

Most 2D problems that we are going to treat are ``near-spherical''
with a moderate departure from a general spherical symmetry.
The radiation field is usually dominated by a central source in these
cases, and it is practical to treat them in a spherical coordinate system.
Therefore, we decided to use $r$, $\beta$, and $\epsilon$ (see Figure~\ref{fig2}
for definition) for reference in our code.
\begin{figure}
\resizebox{\hsize}{!}{
\includegraphics[angle= 270]{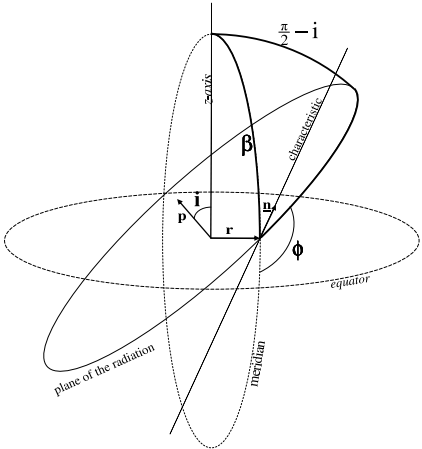}
}
\caption{
Diagram illustrating the connection between the radiation angle $\phi$ and the
inclination angle.
The ``plane of the radiation'' includes the characteristic and the origin.
Angles $i$ and $\beta$ are the angular distances between the $z$-axis and
the directions of the {\bf p} and {\bf r} vectors, respectively.
Eq.~\ref{eq:ct2} can be derived by a spherical sine law using the
boldface spherical triangle.
}
\label{fig3}
\end{figure}

In spherical symmetry, the most natural way to map the directional variation
of the intensity is using the  ``so-called'' radiation coordinates,
$\theta$ and $\phi$, that are defined by
\begin{equation}\label{eq:theta}
cos(\theta)= \underline{\bf n} \cdot \underline{\bf r}
\end{equation}
and
\begin{equation}\label{eq:phi}
 sin(\theta) \cdot sin(\beta) \cdot cos(\phi)= \left[ \underline{\bf n} \times \underline{\bf r}
\right] \cdot \left[ \underline{\bf r} \times \underline{\bf z} \right]   \; .
\end{equation}
The unit vectors $\underline{\bf n}$, $\underline{\bf r}$, and $\underline{\bf z}$ are
pointing in the direction of the radiation, in the radial direction, and in the positive side
of the z axis, respectively (see Fig.~\ref{fig2}).
A proper choice of $\theta$ angle grid can be very useful in treating inherent
discontinuities around the limb of the central star and the symmetries due to the
forward-peaking nature of the radiation field.

As mentioned in \S\ref{section:solution} a serious drawback of the SC method
is the interdependency of the specific intensities at different grid points.
Beside introducing systematic errors by the successive intensity interpolations,
the SC method also couples the directional sampling of the radiation field on
the grid.
Our choice of directions at a grid point not only has to suit the needs of the
particular point but also has to be able to provide suitable starting values
($I_{BC}$) for other points.
Unfortunately, $\theta$ and $\phi$ vary along a characteristic so it is
not possible to use a uniform $\theta$ and $\phi$ grid for all grid points
without intensity interpolations in the radiation coordinates.
The later option is not desirable for multidimensional RT.
First, it requires a large amount of memory to store all intensities for the
interpolation.
Second, it makes the parallelization of the code difficult.

To find a proper directional sampling method one needs to look for quantities that are
conserved along a characteristic, like
\begin{equation}
{\bf p}= {\bf r} \times \underline{\bf n} \; ,
\end{equation}
which we call the ``impact-parameter vector'' (see Fig.~\ref{fig2}).
This vector describes all essential features of a characteristic and can be
considered as an analog of the orbital momentum vector in two body problems.
Its absolute value $p$= $|${\bf p}$|$ is the traditional impact-parameter and its
orientation defines the ``orbital plane'' of the radiation (the plane that contains
the characteristic and the origin).
Following this analogy one can define an ``inclination'' angle for this plane by
\begin{equation}\label{eq:i}
p \cdot cos(i)= {\bf p} \cdot \underline{\bf z} \; .
\end{equation}
In our code we set up a universal grid in impact-parameters ($p$) and in inclination
angles ($i$) for directional sampling.
As opposed to the $\theta$ and $\phi$ angles, the inclination angle and the impact-parameter
do not vary along a ray; therefore, intensities in the proper directions will be available for
the interpolation when the transfer is solved for a given $i$ and $p$.
Using an impact-parameter grid to avoid interpolation in $\theta$ angle has
already been incorporated into previous works \citep[e.g.,][]{bus00}.
By introducing the inclination angle grid we simply exploited the full potential of this
approach.

It is useful to examine the relationship between the radiation angles and our directional
coordinates. The conversion is via
\begin{equation}\label{eq:ct1}
sin( \theta ) = \frac{p}{r}
\end{equation}
and
\begin{equation}\label{eq:ct2}
sin( \phi )= \frac{cos(i)}{sin( \beta )}
\end{equation}
at each grid point.
Equation \ref{eq:ct2} can be easily derived by spherical trigonometry as
illustrated by Fig.~\ref{fig3}.
One can see from Eqs.~\ref{eq:ct1} and  \ref{eq:ct2} that there is a
degeneracy between ``incoming''--``outgoing'', as well as between
``equator-bound''--``pole-bound'' rays.
(The ``pole-bound'' rays are defined by $\frac{\pi}{2} < \phi < \frac{3}{2} \pi$.)
The radiation coordinates ($\theta$, $\phi$) and ($\pi - \theta$,
$\pi - \phi$) are represented by the same ($p$, $i$) pair.
Fortunately, the ``switch-over'' can only occur at certain spatial positions.
For example, the incoming rays become outgoing only at $r$= $p$, so this is just a
simple book-keeping problem.
Nevertheless, one should always bear this degeneracy in mind when doing the actual
programming implementation of our method.

\begin{figure*}
\centering
\includegraphics[width=17cm]{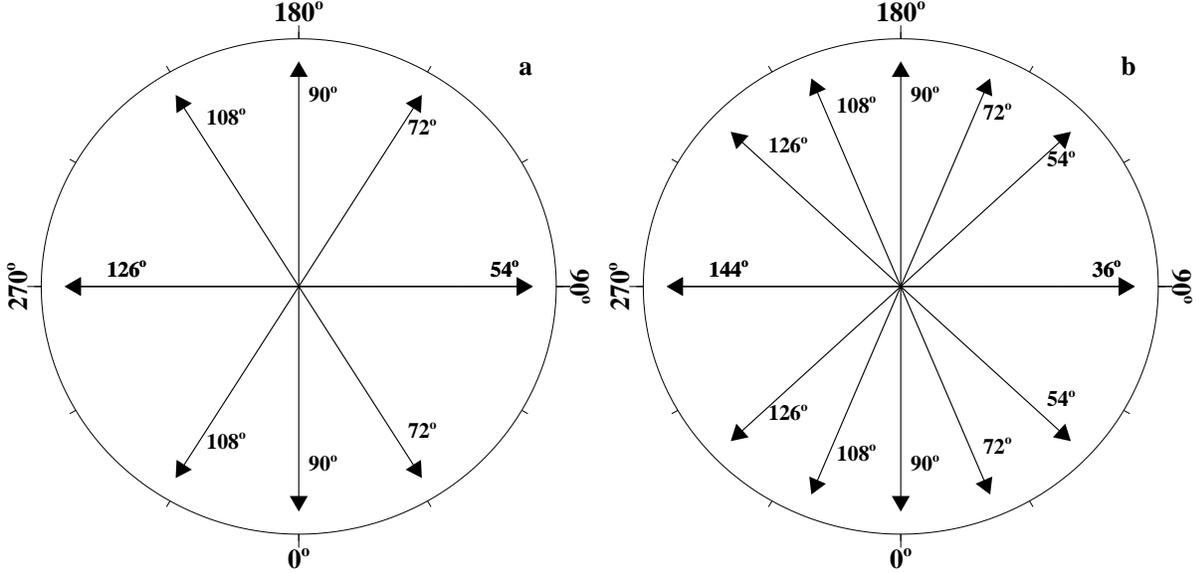}
\caption{
The $\phi$-angle plane at different latitudes as viewed by an observer facing
the central star/object.
The unit vector pointing out of the page is toward the observer.
Each figure is centered on the line of sight of the observer and the equator is
toward the bottom of the page.
The figure was created for inclination angles of 0$^{\rm o}$, 18$^{\rm o}$,
36$^{\rm o}$, 54$^{\rm o}$, 72$^{\rm o}$,  90$^{\rm o}$, 108$^{\rm o}$,
126$^{\rm o}$, 144$^{\rm o}$, 162$^{\rm o}$, and 180$^{\rm o}$ which are
indicated near the head of the arrows.
The radiation angle $\phi$ is measured counter-clockwise from the direction
toward the equator as indicated on the outer rim of the circles.
Panels a and b are for $\beta$= 36$^{\rm o}$ and 54$^{\rm o}$, respectively.
For clarity, we assumed that the impact-parameter ($p$) of
the rays is equal to $r$; therefore, any direction that we sample lies in the
$\phi$ plane.
The figure shows that the $\phi$-angle coverage is latitude dependent and
unevenly spaced.
Note, for example,  the absence of $i$= 0$^{\rm o}$, 18$^{\rm o}$, 36$^{\rm o}$
(and their complementary angles) for $\beta$=  36$^{\rm o}$.
}  \label{fig4}
\end{figure*}

There remains one important question.
How exactly do we choose the actual impact-parameter and inclination angle
grid?
We adopted the approach of \cite{bus00} who used the radial grid and a number of
``core rays'' ($p \leq r_{core}$) for the impact-parameters.
The core rays are added only if a central source with a radius $r_{core}$ is present
in the model.
This will provide a radius dependent sampling since only $p \leq r$ can be used for a
given $r$ radius.
Also, the sampling is uneven and sparser around $\theta$= $\frac{\pi}{2}$ than around
$\theta$= 0 or $\pi$.
Nevertheless, this grid was proven to be adequate for near spherical problems and
also very convenient to use.
For example, it ensures that $p$= $r$ (the switch-over from ``incoming'' to ``outgoing''
ray) is always a grid point.
Similarly, we based our inclination angle grid on the $\beta$ grid, although, we
have the option to define it independently.
If needed, extra inclination angles can also be included around $i$=
$\frac{\pi}{2}$ to increase the $\phi$ angle resolution at higher latitude.

Fig.~\ref{fig4} illustrates a typical inclination angle grid and the $\phi$-angle
sampling it provides.
For illustration purposes we use a hypothetical $\beta$ grid of $\frac{1}{2} \pi$ (equator),
$\frac{8}{10} \pi$ (72$^{\rm o}$), $\frac{6}{10} \pi$ (54$^{\rm o}$), $\frac{4}{10}
\pi$ (36$^{\rm o}$), $\frac{2}{10} \pi$ (18$^{\rm o}$), and 0 (pole).
Then, one may choose these $\beta$ values and their corresponding complementary
angles ($\pi$-$\beta$) for the inclination angle grid.
By our definition, angles $i \leq \frac{\pi}{2}$ sample the 0~$\leq$~$\phi$~$\leq$~$\pi$ range,
while $i > \frac{\pi}{2}$ covers the rest of the $\phi$ space (see Fig.~\ref{fig4}).
The behavior of the $\phi$-angle sampling created by this inclination angle grid is very
similar to that of the $\theta$-angle sampling provided by the radial grid.
One can easily see from Eq.~\ref{eq:ct2} that for a given $\beta$
any $i < \frac{\pi}{2} - \beta$ has no solution for $\phi$.
The equatorial regions ($\beta \sim \frac{\pi}{2}$), therefore, are well sampled in
$\phi$ angle while there is only one valid inclination angle at $\beta$= 0 ($i$=
$\frac{\pi}{2}$).
This is reasonable in axi-symmetrical models, as long as the polar direction is also
the axis of symmetry (as we explicitly assume).
The $\phi$-angle sampling is also uneven.
The regions around $\phi$= 0 and $\pi$ (local meridian) are better resolved than those
around $\phi$= $\frac{\pi}{2}$ and $\phi$= $\frac{3}{2} \pi$.
In \S\ref{section:2Dstat}--\ref{section:2Dwind} we will demonstrate
that our sampling method not only eliminates the need for interpolations in $\theta$ and
$\phi$ angles, but sufficiently recovers the directional variation of the radiation at every point
and is adequate for RT calculations in axi-symmetric envelopes.

\subsection{Approximate Lambda Iteration}\label{section:ALO}

A seemingly natural choice for the iteration between the RT and level
populations is the notorious ``$\Lambda$-iteration''.
In this scheme, the level populations from the previous cycle are used
to calculate new $J$ and $\overline{J}_l$ which in turn are used to update
the populations.
Unfortunately, this simple procedure fails to converge for large optical depths.
Convergence is ensured, however, by using the Accelerated Lambda Iteration
\citep[ALI; see e.g.,][]{ryb91, hub92} which takes some of the inherent coupling
into account implicitly.
The relationship between $J$  and the source function $S$ can be summarized as
\begin{equation}\label{eq:LA}
J = \Lambda \left[ S \right] \; ,
\end{equation}
where the $\Lambda$ operator can be derived from Eqs.~\ref{eq:J} and \ref{eq:RT}.
Both $\Lambda$ operator and $S$ depend on the level populations, however,
we can ``precondition'' $\Lambda$ \citep[i.e., use the populations from the
previous
cycle to evaluate it, see e.g.,][]{ryb91} and only take the coupling through $S$ into
account to accelerate the iteration.
In 2D, $\Lambda$ in its entirety is too complicated to construct and
time consuming to invert, which is necessary to take the coupling into account.
We can, however, split the $\Lambda$ operator into an ``easy-to-invert'' $\Lambda^*$
(Approximate Lambda Operator) and the remaining ``difficult'' part by
\begin{equation}\label{eq:ALO}
J = \Lambda^* \left[ S \right] \; +  \left( \Lambda - \Lambda^*  \right)
\left[ S \right] \; .
\end{equation}
Then, we can precondition the ``difficult'' part by using the old populations, and
accelerate the iteration by inverting $\Lambda^*$.
Note, that the full $\Lambda$ operator never needs to be constructed, only
$\Lambda^*$ since
\begin{equation}\label{eq:ALO2}
\left( \Lambda - \Lambda^*  \right) \left[ S^{i-1} \right]= J^{i-1} \; - \;  \Lambda^*
\left[ S^{i-1} \right]
\end{equation}
where $J^{i-1}$ and $S^{i-1}$ is the moment and source term from the previous
iteration cycle.

The actual form of  $\Lambda^*$ is a matter of choice as long as it can be easily
inverted.
The most practical in 2D is separating out the local contribution
(i.e., diagonal part of the $\Lambda$ operator when written in a matrix form).
This is easy to calculate and has reasonably good convergence characteristics.
During the evaluation of moments $J$ and $J_l$ (see \S\ref{section:solution}),
we also calculate the diagonal $\Lambda^*$ operator.
This is a fairly straightforward book-keeping since we just have to add up the
weights used for the local source function during the
integration of Eq.~\ref{eq:I}.

We used the $\Lambda^*$ operator to accelerate the ES
iterations in our test calculations (see the following sections).
Apart from the initial ``hiccups'' of code development, the operator always
worked as expected and produced the published convergence
characteristics \citep{ryb91}.
The implementation of the ALO iteration into the solution of the statistical
equilibrium equation is discussed in {\bf \citetalias{geo05}}.

\section{Code Verification and Test Results} \label{section:tests}

We have developed a C++ code that implements the solution technique
described in \S\ref{section:code}.
As mentioned in \S\ref{section:solution}, we used a modified version of the
traditional SC method by terminating the characteristics at the closest spherical
shell rather than any cell boundary (i.e., our SCs cross cell boundaries in $\beta$
direction).
This modification allows us to avoid intensity interpolations in the radial direction
which increases the accuracy when a strong central source dominates the radiation field.
The transfer calculation for an impact-parameter ($p$) and inclination angle ($i$)
pair is performed on an axi-symmetric torus with an opening-angle of 2$i$ and which is
truncated at the inner radius of $r$= max($p$, $r_{core}$).
This torus contains all spatial regions that a ray described by $p$ and $i$ can reach.
The calculation starts at the outermost radius and proceeds inward, shell by shell, until
the truncation radius is reached; then, the outgoing radiation is calculated in a similar
manner by proceeding outward.
At the outer boundary we set the incoming intensity to zero while either a diffusion
approximation or a Schuster-type boundary condition can be used at the truncation
radius if it is equal to $r_{core}$.
In its present form, the code assumes top-bottom symmetry, however, this
approximation can easily be relaxed to accommodate general axi-symmetric models.
The RT calculation for each ($p$, $i$) pair is independent from any other.
The only information they share are the hydrodynamic structure of the envelope,
the opacities, and emissivities; all of which can be provided by ASTAROTH.

There are at least two major venues to accommodate multi-processor calculations
in the code.
One way is to distribute the ($p$, $i$) pairs among the available processors.
To optimize the calculation one needs to resolve a non-trivial load-sharing issue.
The actual number of spatial grid points involved in the RT is not the same for all
($p$, $i$) pairs, so the duration of these calculations is not uniform.
For example, the transfer for $p$= 0 and $i$= $\frac{\pi}{2}$ involves all spatial
grid points, while the one for  $p$= 0 and $i$= $0$ includes only the points lying
in the equator.
To use the full capacity of all processors at all times, a proper distribution mechanism
needs to be developed that allows for the differences between processors and the
differences between ($p$, $i$) pairs.

We also have the option to distribute the work among the processors by distributing
the frequencies for which the RT is calculated.
In this case, the work-load scales linearly with the number of frequencies, so the
distribution is straightforward.
However, the lack of sufficient memory may prevent the distribution of all
opacities and emissivities and the processors may have information only over their
own frequency range.
To take the effects of velocity field into account at the limiting frequencies, we
introduce overlaps between the frequency regions.

So far, we have performed multi-machine calculations where the ($p$, $i$) pairs or
frequency ranges were distributed by hand.
The results of the distributed calculations were identical to those performed on a
single machine.
Work is under way to fully implement distributed calculations by using MPI protocols.
Since our goal is to run the entire stellar atmosphere code on multiple
processors, we will discuss the details of parallelization in a subsequent
paper after we have fully integrated our code into ASTAROTH.

In the following we describe the results of some basic tests of our code.
First, we calculate the radiation field in static 2D problems with and without ES.
Then, we present our results for realistic spherical problems with substantial wind
velocities.
Finally, we introduce rotation in a spherical model and demonstrate
the ability of our code to handle 2D velocity fields.

\subsection{Static 2D Models}\label{section:2Dstat}

The basic characteristics of our code  were tested by performing simple calculations,
1D and 2D models without velocity field.
We used the results of a LC program developed by {\cite{hil94, hil96}} as a benchmark.
This code was extensively tested and verified by reproducing one dimensional models as
well as analytical solutions available for optically thin stellar envelopes {\citep[e.g.,][]{bro77}}.
It was also tested against Monte-Carlo simulations of more complicated models.

\begin{figure}
\resizebox{\hsize}{!}{
\includegraphics[angle= 270]{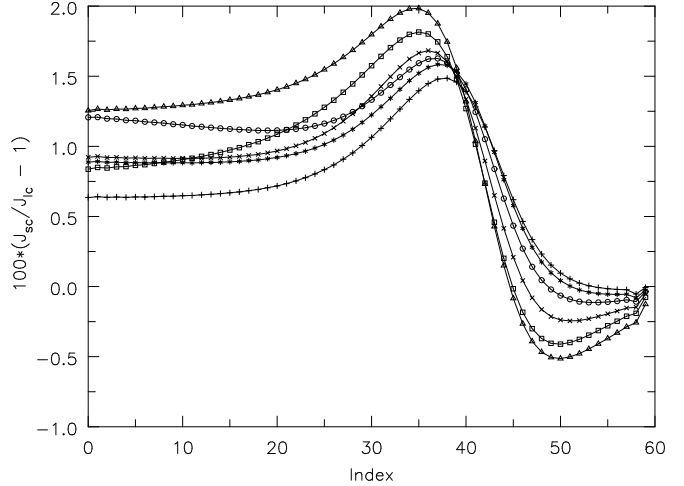}
}
\caption{
The percentage difference between the J moments calculated by our (J$_{sc}$) and
by the LC program (J$_{lc}$) as a function of the depth
index (0 and 59 are the indices of the outer-most and inner-most radial grid
points, respectively) for different latitudes.
The ES opacity in this model is described by Eq.~\ref{eq:ES}.
The symbols +, *, o, x, $\Box$, and $\triangle$ indicate the differences
for $\beta$= 0, 0.1$\pi$, 0.2$\pi$, 0.3$\pi$, 0.4$\pi$, and $\frac{\pi}{2}$, respectively.
Our code systematically overestimates $J$ in the outer regions
(0--40) which is mostly due to the second order accuracy of the radial interpolations.
Errors from other sources (e.g., latitudinal resolution, $\phi$ angle sampling) are
most important at high-latitudes ($\beta \sim$ 0.1--0.2 $\pi$) but still contribute
less than $\sim$1\%.
}
\label{fig5}
\end{figure}

\begin{figure}
\resizebox{\hsize}{!}{
\includegraphics[angle= 270]{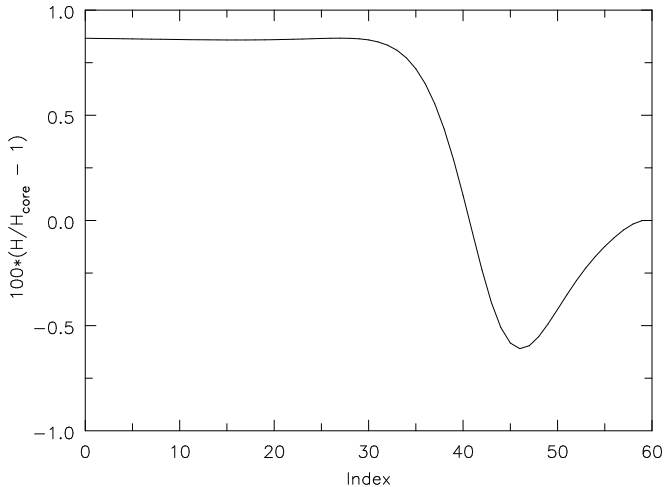}
}
\caption{
The percentage loss/gain of the total radial flux ($H= \int_{4 \pi} H_r r^2 d
\Omega$) with respect to the total flux emanating from the stellar surface
(H$_{core}$) as a function of depth index for the 2D model described
by Eq.~\ref{eq:ES}.
The flux is conserved within $\sim$1\%.
}
\label{fig6}
\end{figure}

Our code reproduced the results of the LC program within a few percent for all
spherical and axi-symmetric models.
It was proven to be very stable and was able to handle extreme cases with
large optical depths.
The most stringent tests were the transfer calculations in purely scattering
atmospheres.
In such cases, the necessary iterations accumulate the systematic errors which
highlights any weakness in the program.
Several 1D and 2D scattering models were run with ES optical depths varying between
1 and 100.
Figures~\ref{fig5} and \ref{fig6} compare our results to those of the LC code
for a model with electron scattering opacity distribution of
\begin{equation}\label{eq:ES}
\chi_{es} = 10 \cdot \left[ \frac{r_{core}}{r} \right] ^3 \cdot \left( 1 - \frac{1}{2}
\cdot cos^2 \beta \right) \; .
\end{equation}
No other source of opacity and emissivity was present in the model.
At the stellar surface we employed a Schuster-type  boundary condition of
$I_{BC}$ = 1, while $I_{BC}$= 0 was used at the outer boundary.
The ES iteration was terminated when $\frac{\Delta J}{J} \le$~0.001\%
had been achieved.
This model is an ideal test case since the ES optical depth is large enough to require a
substantial number of iterations to converge, but the convergence is fast enough to allow
for experimenting with different spatial resolutions.

For the results we present in Fig~\ref{fig5}, the LC code was run with 60
radial and 11 latitudinal grid points.
The $\phi$ radiation angle was sampled in 11 directions evenly distributed
between 0 and $\pi$.
This code assumes top-bottom and left-right symmetry around the equator
($\beta$= $\frac{\pi}{2}$) and the local meridian ($\phi$= 0), respectively,
so only half of the $\beta$ and $\phi$ space had to be sampled.
The radial grid, supplemented by 14 core rays, was used to map the $\theta$
radiation angle dependence (see \S\ref{section:dir} for description).
We used a slightly modified radial and latitudinal grid in our code.
We added 3 extra radial points between the 2 innermost depths of the original
grid, and 6 extra latitudinal points were placed between $\beta$= 0 and
0.15~$\pi$.
These modifications substantially improved the transfer calculation deep
in the atmosphere and at high latitudes.
The sampling method of the $\theta$ angle was identical to that of
the LC code.
We based our inclination angle grid on the $\beta$ grid and added 4 extra
inclination angles around $\frac{\pi}{2}$ to improve the coverage at high latitudes.
This grid resulted in a latitude dependent $\phi$ angle sampling.
At the pole, the radiation was sampled in only 2 directions while on the
equator 60 angles between 0 and $2 \pi$ were used.
Note, that our code does not assume left-right symmetry!

Figures \ref{fig5} and \ref{fig6} show that  we were able to reproduce the results
of the LC code within $\sim$2\% accuracy, and the total radial flux is conserved
within 1\% level.
It is also obvious that our code needs higher spatial resolution to achieve the
accuracy of the LC code.
This is expected since the LC program uses higher order approximations and adds
extra spatial points when needed to increase the overall accuracy.
In fact, one should not call the LC code a pure $N_r$= 60, $N_{\beta}$= 11 model.
The auxiliary points increased the real resolution.
It is not surprising, on the other hand, that our code runs substantially faster on
the same machine.
The difference was between a factor of 10 and 2, depending on the number of
iterations needed to converge.
Unfortunately, we have not yet introduced sophisticated acceleration techniques,
like the Ng acceleration \citep{ng74}, so our code is not the most efficient when
a very large number of iterations is needed.

The agreement between our code and the LC code progressively worsened
as the total ES optical depth increased.
Satisfactory agreement could be achieved, however, by increasing
the radial resolution.
Our test problems and most of the real problems that we will address later
are near spherical with a modest latitudinal variation.
The intensity  reflects the strong radial dependency and, therefore,
the radial resolution controls the overall accuracy.
Fig.~\ref{fig5} reveals another feature of our method that affects the
accuracy.
Our result is sensitive to the high-latitude behavior of the intensity
for a given inclination angle and impact-parameter.
At the high-latitude regions, a given inclination angle samples directions
that can be almost parallel with the equator.
Slightly different directions that are almost parallel with the equator can sample
very different radiation in some axi-symmetric models, such as models with
thin disks.
Aggravating this problem, our method also uses fewer directions to map the
radiation field at these high latitudes, unless extra inclination angles around
$\frac{\pi}{2}$ are included.
This explains why we had to use extra latitudes and inclination angles
to produce the result for Figs.~\ref{fig5} and  \ref{fig6}.
We would like to emphasize, however, that these problems are important
only in extreme axi-symmetric models (e.g, very thin disks or strong polar
jets).
Many times, as it will be demonstrated in the next sections, reasonable
accuracies can be achieved on ordinary and simple grids.

During the static 2D tests, we also experimented with the multi-grid capability
of our code and verified its scaling behavior.
Tests with progressively increasing spatial resolution showed that our code has
second order accuracy.
By doubling the number of radial grid points, for example, the errors decreased
roughly 4-fold.
We also performed the ES iterations in multiple steps and at progressively
increasing resolution.
First, a coarse grid was created (e.g., half of the nominal resolution) for a crude and
fast initial iteration.
Then, with the updated source terms, a second iteration was performed
on the nominal grid.
This ``double iteration'' scheme was generally a factor of two faster than a single
iteration on the nominal grid.
This approach will be a promising venue for fast iterations in combination
with other acceleration techniques.

\subsection{1D Test Cases with Realistic Wind Velocities} \label{section:1Dwind}

After performing static 2D tests, we applied our code to realistic 1D atmospheres.
The primary purpose of these tests was to verify our handling of realistic velocity
fields.
We used a well known and tested 1D stellar atmosphere code,  CMFGEN \citep{hil98},
for comparison.
Observed spectra for a CMFGEN model are calculated independently by an
auxiliary routine, CMF\_FLUX \citep[see][for a description]{bus05}.
We compared our simulated observed spectra to those of CMF\_FLUX.

\begin{table}
\caption{Description of Model v34\_36C\label{tab1}}
\begin{tabular}{llr} \hline\hline
Star               &~~~~~~~~ &  \object{AV~83}  \\
Sp.~Type      &~~~~~~~~ &  O7~Iaf        \\
log~g            &~~~~~~~~ &  3.25        \\
R                  &~~~~~~~~ & 19.6 R$_{\odot}$ \\
T$_{eff}$     &~~~~~~~~ & 34000~K \\
\.{M}            &~~~~~~~~ & 2.5$\times$10$^{-6}$~M$_{\odot}$~yr$^{-1}$ \\
V$_{\infty}$ &~~~~~~~~ & 900~km~s$^{-1}$ \\
$\beta^a$      &~~~~~~~~ & 2 \\ \hline
\end{tabular}

$^a$ -- Power for CAK velocity law \citep{CAK}.
\end{table}
We have an extensive library of CMFGEN models to choose
a benchmark for our tests.
We picked \object{AV~83}, a supergiant in the SMC (see Table~\ref{tab1}) which
was involved in a recent study of O stars \citep{hil03}.
Accurate rotationally broadened spectra with different viewing
angles are also available for this star \citep{bus05} which we will use
for comparison in \S\ref{section:2Dwind}.
A detailed description of the CMFGEN models for \object{AV~83} can be found in
\cite{hil03}.
We chose their model v34\_36C (see Table~\ref{tab1}) to test our code.
The radial grid with 52 depth points was adopted from this model.
The impact-parameter grid which samples the $\theta$ radiation angle
was defined by the radial grid augmented by 15 core rays (see \S\ref{section:dir}
for details).
Our simulation was run as a real 2D case with two latitudinal angles ($\beta$= 0 and
 $\frac{\pi}{2}$).
We used 3 inclination angles which resulted in transfer calculations
for 2 and 4 $\phi$ angles in the polar and the equatorial directions, respectively.
The RT calculations were performed on frequency regions centered around
strategic lines, like H$\alpha$.
A coarse grid ($N_r$= 26, $N_{\beta}$= 2) and the nominal ($N_r$= 52, $N_{\beta}$= 2) grid was
used for the ES iteration as in the cases of static models (see \S\ref{section:2Dstat}).
Note, that our model is not a fully consistent solution because we
did not solve for the level populations.
We simply used the output opacities and emissivities of the converged
CMFGEN model and calculated the RT for it.
\begin{figure*}
\centering
\includegraphics[width=16cm]{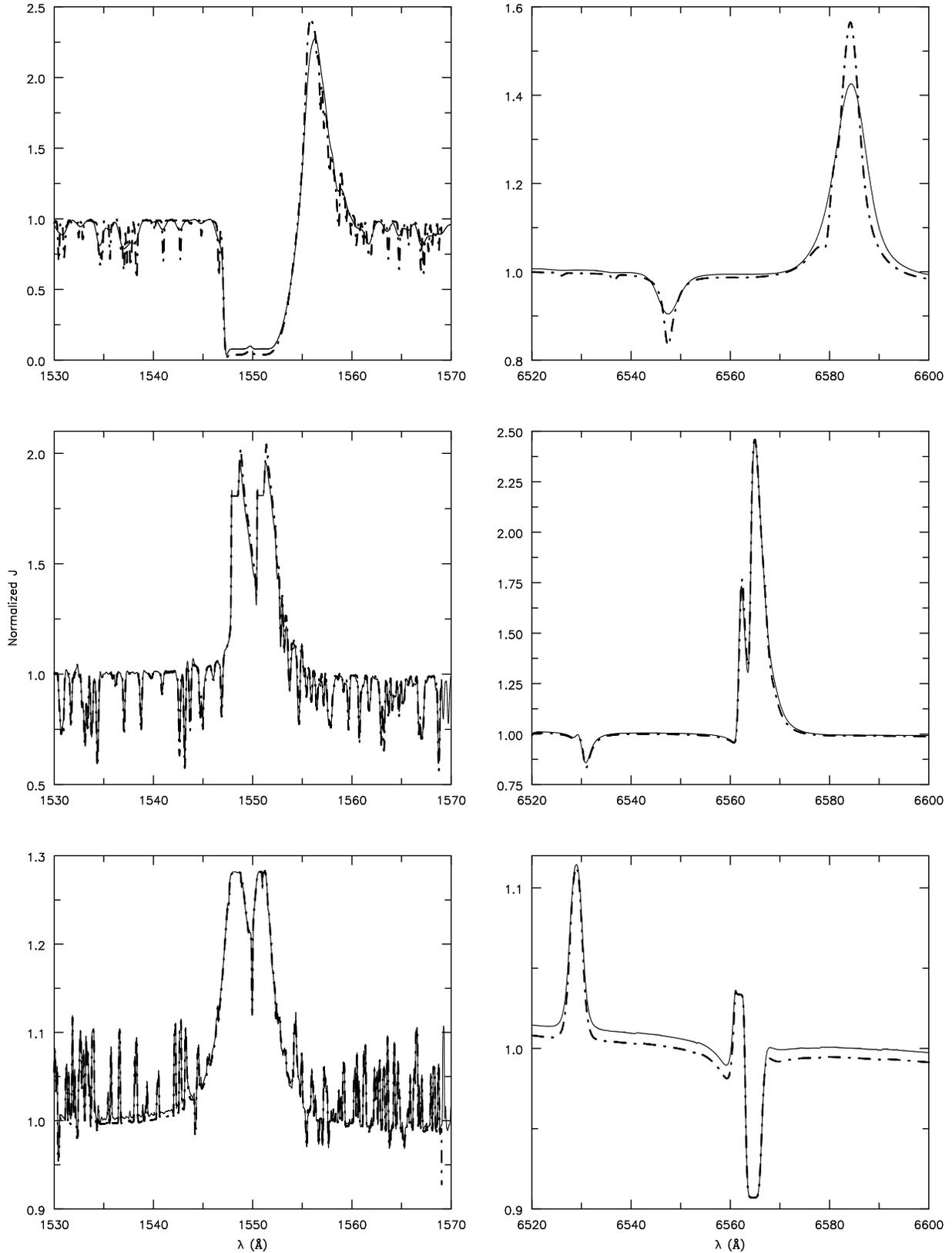}
\caption{
The normalized $J$ moment as a function of wavelength around the \ion{C}{IV}
$\lambda\lambda$1548--1552 doublet (left column) and $H\alpha$ (right column)
at different locations in the envelope of \object{AV~83} (the stellar model is described in
Table~\ref{tab1}).
The top row of figures shows $J$ at $v_r \sim v_{\infty}$, the middle at
$v_r \sim 0.1 v_{\infty}$, while the bottom row displays $J$ in the hydrostatic
atmosphere ($v_r \sim 0$).
Note, that all spectra are in the co-moving frame.
The solid (thin) and dash-dotted (thick) lines were calculated by
CMFGEN and our code, respectively.
Even though this model is spherical, our code treated it as a 2D case.
As expected for spherical models, we calculated identical $J$ moments for
every latitude.
}
\label{fig7}
\end{figure*}
\begin{figure*}
\centering
\includegraphics[width=17cm]{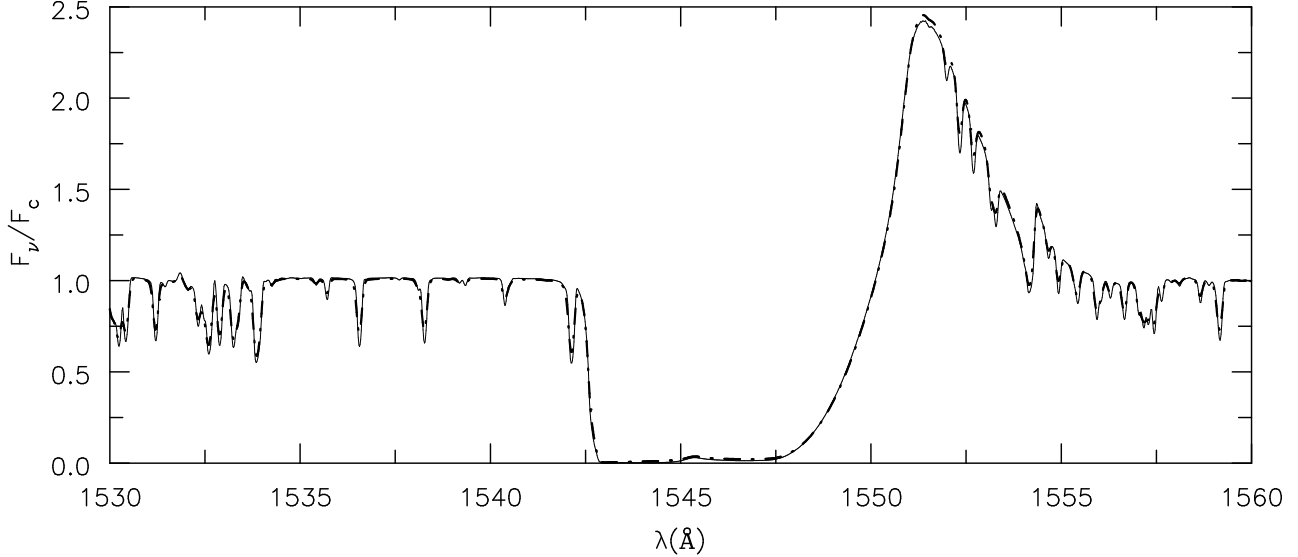}
\caption{
The observed spectrum around the \ion{C}{IV}  $\lambda\lambda$1548--1552
doublet calculated by CMF\_FLUX (thin solid line) and by our code (thick dash-dotted
line).
Note, that these spectra are in the observers' frame.
}
\label{fig8}
\end{figure*}

Figure~\ref{fig7} shows the normalized $J$ moment as a function of wavelength
for the \ion{C}{IV} $\lambda\lambda$1548--1552 doublet and H$\alpha$ at
different depths.
The results of our code and those of CMFGEN are in good agreement, except that
we resolve narrow lines better.
CMFGEN solves the moment equation in the co-moving frame, starting at the
largest frequency.
This procedure introduces bleeding which broadens the sharp lines.
Our results are not affected by this bleeding since we use the
formal solution.

Figure~\ref{fig8} shows the observed spectrum for \object{AV~83} in the observer's
frame. As is the case with the $J$ moment the agreement between our
code and CMF\_FLUX is excellent.
In this case CMF\_FLUX does a better job but this is expected.
Our code is primarily for providing $J$ and $\overline{J}_l$ for the solution of the
rate equations while it produces observed spectra only for testing.
The main purpose of CMF\_FLUX, on the other hand, is to produce highly accurate
spectra in the observer's frame.

We would like to emphasize that our code did not need higher spatial resolution to
reproduce the results of CMFGEN/CMF\_FLUX, as opposed to some cases
presented in \S\ref{section:2Dstat}.
The pure scattering models of \S\ref{section:2Dstat} were extreme examples
and were hard to reproduce.
The comparison with CMFGEN proves that our code can handle realistic problems at a
reasonable spatial resolution.

\subsection{Tests with a Rotating Envelope}\label{section:2Dwind}

As a final test for our SC code we ran simulations of semi-realistic
2D atmospheres.
These were created by introducing rotation in otherwise 1D models.
\object{AV~83} offers a good opportunity for such an experiment.
It has a slowly accelerating wind and low terminal velocity that enhances
the importance of the rotational velocities.
Also, its spectrum contains numerous photospheric and wind features
which behave differently in the presence of rotation.

Capitalizing on these features \cite{bus05} used \object{AV~83} to test their code
for calculating observed spectra in 2D models, and to perform a comprehensive
study of the observable rotation effects.
They utilized the LC method and a very dense directional sampling to calculate
the observed spectra for an arbitrary viewing angle.
This code serves the same purpose for ASTAROTH as CMF\_FLUX  does for
CMFGEN; to calculate
very accurate observed spectra for an already converged model.
Since our code produces observed spectra only for testing purposes and error
assessment, the comparison provides only a consistency check between
the two codes.
Further, \cite{bus05} do not calculate radiation moments, so we could only examine
whether our results behave as expected with respect to the 1D moments of CMFGEN.

The rotation in the envelope of \object{AV~83} was introduced by using the Wind
Compressed Disk model \citep[WCD,][]{bjo93}.
\cite{bus05} ran several calculations to study the different aspects of rotation.
We adopted only those that were used to study the Resonance Zone Effects
\citep[RZE,][]{pet96}.
To isolate RZE-s, the latitudinal velocities were set to zero and the density was
left unaffected by the rotation (i.e., it was spherical).
The azimuthal velocity in such simplified WCD cases is described by
\begin{equation}\label{eq:Vphi}
v_{\phi}= v_{eq} \cdot \frac{r_{core}}{r} \cdot sin \left( \beta \right)
\end{equation}
\citep[see;][]{bjo93, bus05}.
For the maximum rotational speed on the stellar surface ($v_{eq}$) we adopted
250~km~s$^{-1}$ following \cite{bus05}.
The radial velocity in the WCD theory is described by a CAK velocity law, so we
used the same radial velocities as in \S\ref{section:1Dwind}.

We again adopted the radial grid of model v36\_34C \citep{hil03} and
used three $\beta$ angles (0, $\frac{\pi}{4}$, and $\frac{\pi}{2}$).
In addition to these grids we had a dense radial and latitudinal grid ($N_r$= 205,
$N_{\beta}$= 9) for the interpolation of opacities, emissivities, and velocities; and
a coarse grid ($N_r$= 26, $N_{\beta}$= 2) for the ES iteration.
We used 14 inclination angles, evenly spaced between 0 and $\pi$, which
resulted in intensity calculations for 24 $\phi$ angles (between 0 and 2$\pi$) at every
point on the equator.
As before, we performed our ``double ES iteration scheme'' (see
\S\S\ref{section:2Dstat} and \ref{section:1Dwind}) with convergence
criteria of $\frac{\Delta J}{J} \leq$~0.001\%.

\begin{figure*}
\centering
\includegraphics[width=17cm]{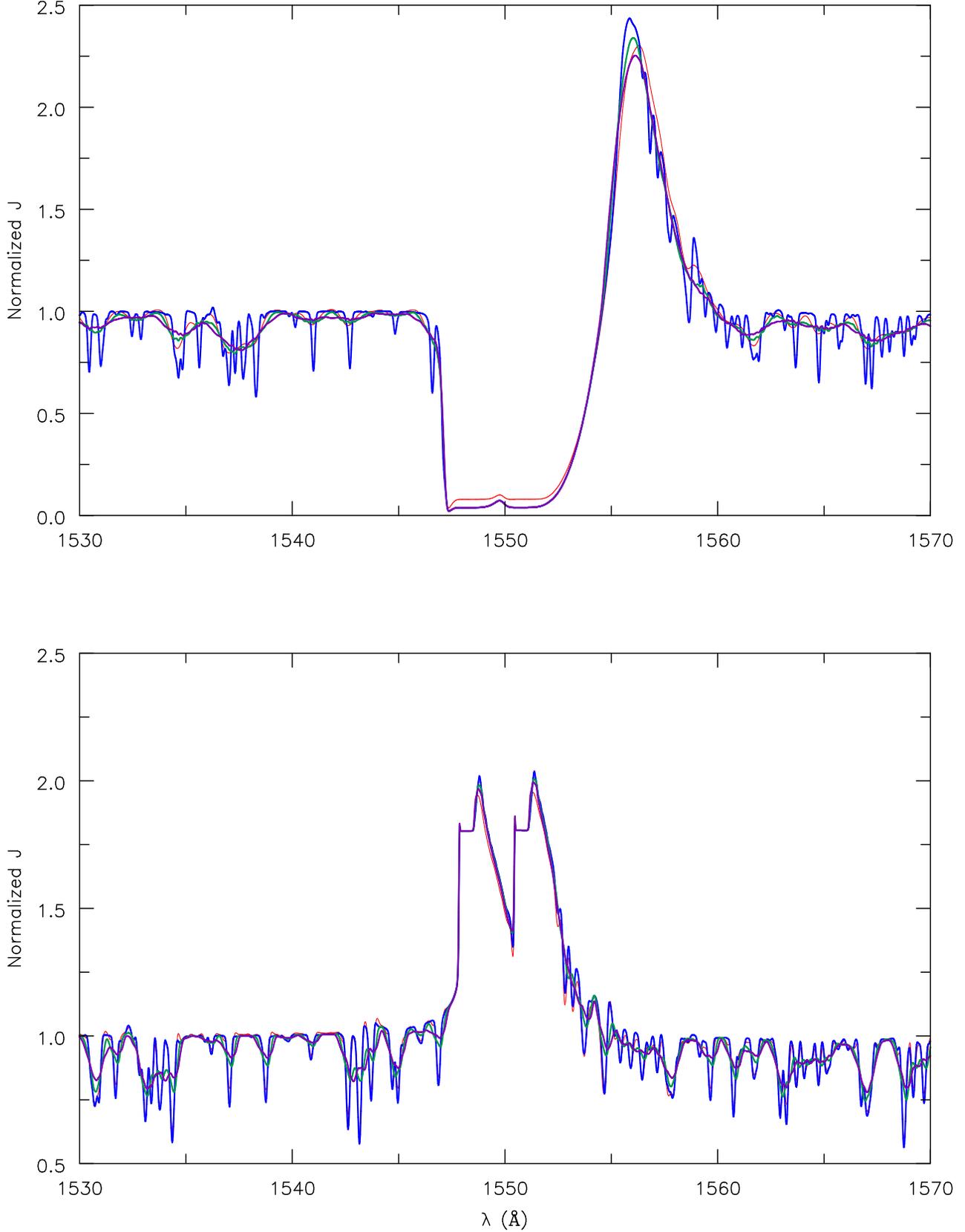}
\caption{
The normalized $J$ moment as a function of wavelength around the \ion{C}{IV}
$\lambda\lambda$1548--1552 doublet at $v_r \sim v_{\infty}$ (top) and at
$v_r \sim 0.1 v_{\infty}$ (bottom).
The wind velocity is described by a simplified version of the WCD
model, for which the polar velocities and the density enhancements were turned off
(see text for description).
The azimuthal rotation was calculated by Eq.~\ref{eq:Vphi} with $v_{eq}$=
250 km~s$^{-1}$.
The thin (red) curve is the basic spherical symmetric model of \object{AV~83} which was
produced by CMF\_FLUX.
The thick blue, green, and purple lines were calculated by our code
and display $J$ for $\beta$= 0, $\frac{\pi}{4}$, and $\frac{\pi}{2}$, respectively.
}
\label{fig9}
\end{figure*}
\begin{figure*}
\centering
\includegraphics[width=17cm]{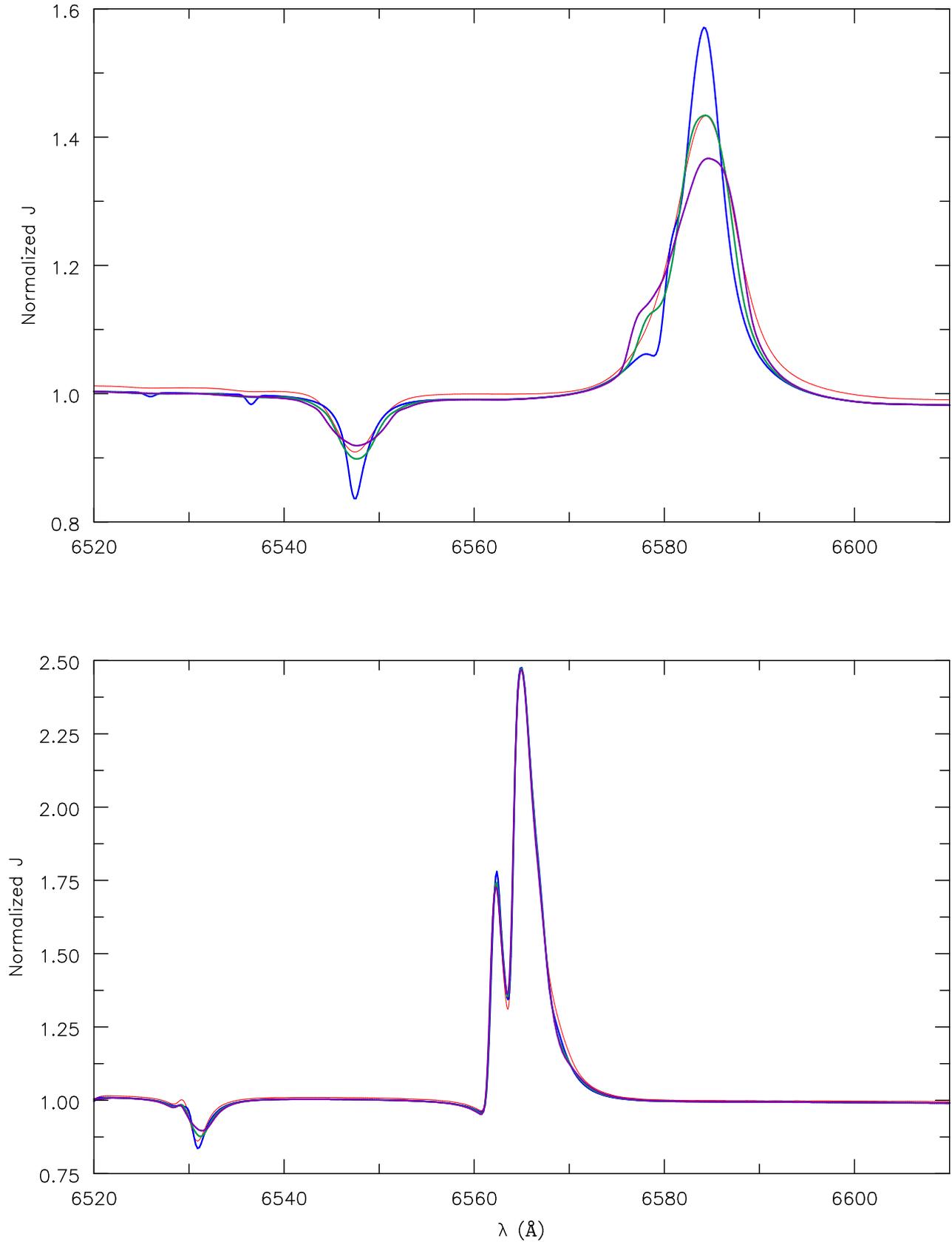}
\caption{
Same as figure~\ref{fig9}, but for the spectra around $H\alpha$.
}
\label{fig10}
\end{figure*}
\begin{figure*}
\centering
\includegraphics[width=16cm]{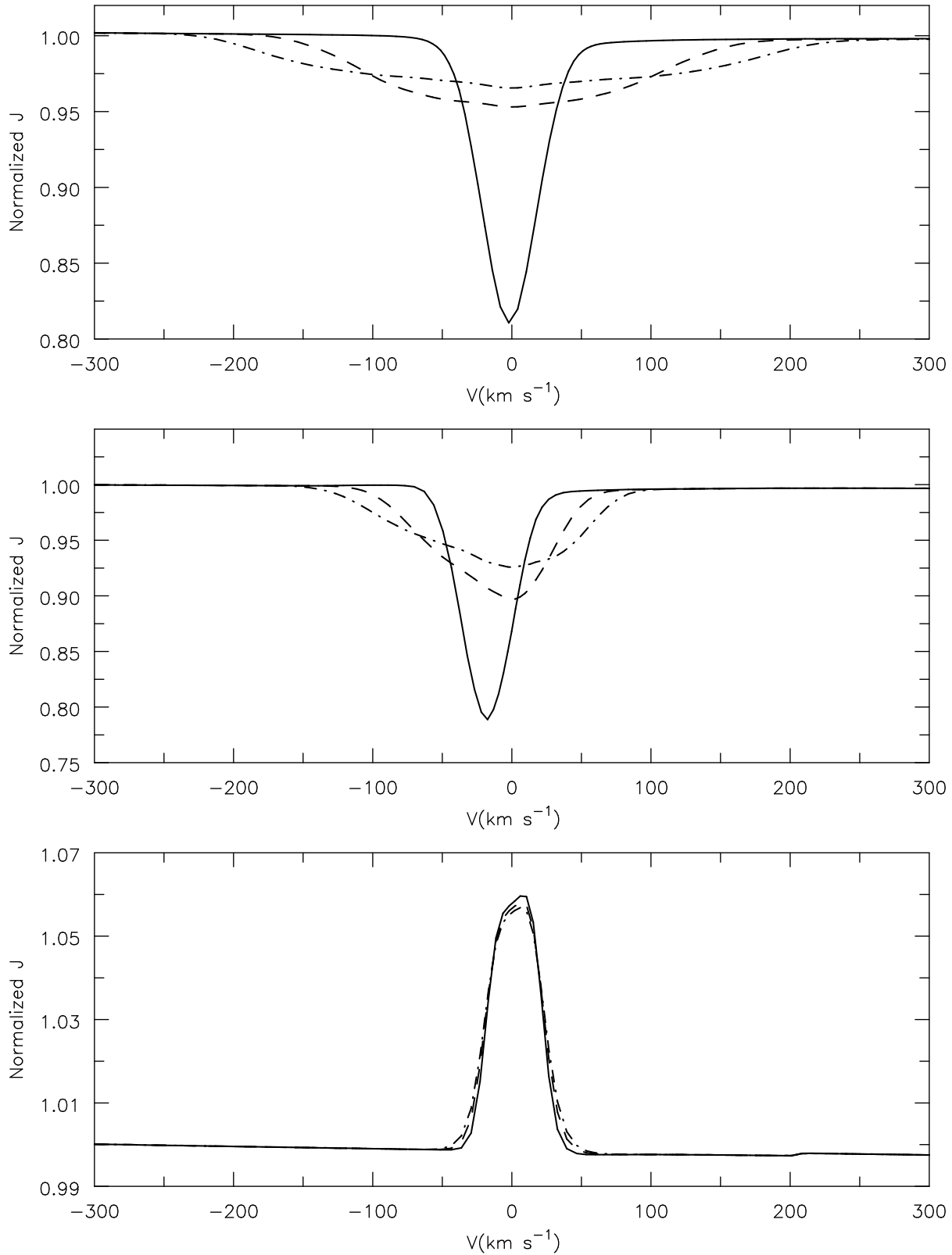}
\caption{
Normalized $J$ profiles of \ion{He}{I} $\lambda$4713.17 at $v_r \sim v_{\infty}$ (top),
$0.1 v_{\infty}$ (middle), and 0 (bottom).
The solid, dashed, and dash-dotted spectra are for $\beta$= 0 (pole), $\frac{\pi}{4}$,
and $\frac{\pi}{2}$ (equator), respectively.
The velocity scale is centered on the line and corrected for the above radial velocities.
Our code reproduces the expected characteristics of the profiles within the
uncertainties of our calculations ($\sim$ 20 km~s$^{-1}$).
Note the skewed line profiles at intermediate radii (middle panel) which are the results
of the broken forward-backward symmetry around the rotational axis
(see \S\ref{section:2Dwind} for details).}
\label{fig11}
\end{figure*}
\begin{figure*}
\centering
\includegraphics[width=16cm]{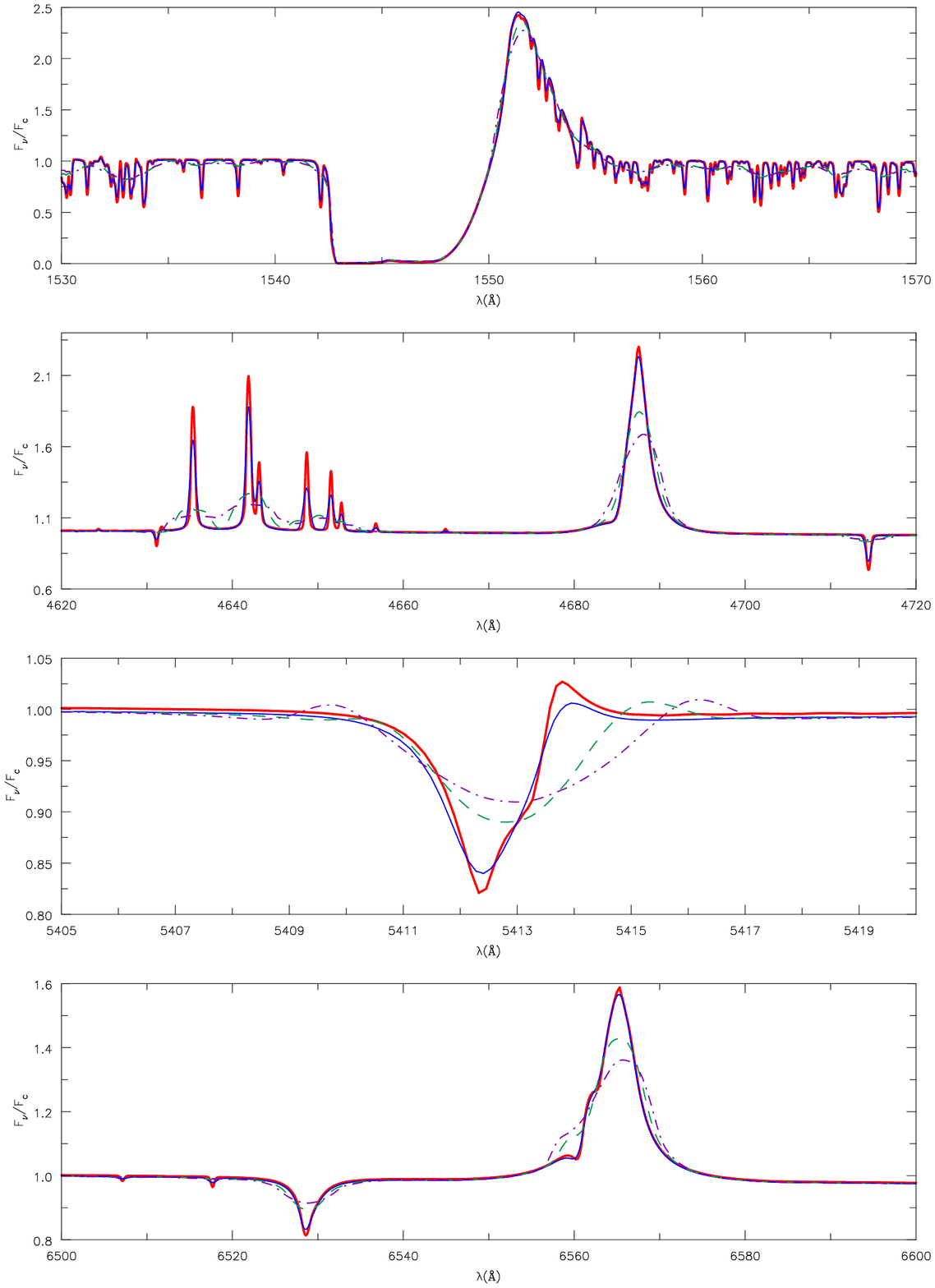}
\caption{
The observed spectra of \object{AV~83} around the \ion{C}{IV}  $\lambda\lambda$1548--1552
doublet (top), the \ion{C}{III}/\ion{N}{III}/\ion{He}{II} emission complex between
4630--4700 \AA ,  \ion{He}{II} $\lambda$5411, and H$\alpha$ (bottom), respectively.
See \S\ref{section:2Dwind} and Fig.~\ref{fig9} for the description of the model
parameters.
The thick (red) curve is the spherical model calculated by CMF\_FLUX, while the
thin (blue), dashed (green), and dashed-dotted (purple) curves are our calculations
for viewing angles 0, $\frac{\pi}{4}$, and $\frac{\pi}{2}$, respectively.
Note that the characteristics of these spectra (e.g., line widths and shapes) are very
similar to those of \citet[see text for further details]{bus05}.
}
\label{fig12}
\end{figure*}

Figures~\ref{fig9} and \ref{fig10} show the behavior of the $J$ moment around the
\ion{C}{IV}~$\lambda\lambda$1548--1552 doublet and H$\alpha$, respectively,
and also for the closest spherical and non-rotating CMFGEN model (thin/red line).
It is obvious that substantial deviation occurs only in the outer envelope and only for
photospheric lines.
Strong P-Cygni profiles, like those of the \ion{C}{IV}~$\lambda\lambda$1548--1552
doublet, are barely affected apart from a little smoothing around the blue absorption edge
and at the maximum emission.
The H$\alpha$ emission, on the other hand, changes its strength substantially between
$\beta$= 0 and $\frac{\pi}{2}$.
This sensitivity casts doubts about the reliability of H$\alpha$ as an accurate mass loss
indicator for rotating stars with unknown viewing angle.
A similar sensitivity to the rotation can also be seen on the iron lines around
\ion{C}{IV}~$\lambda\lambda$1548--1552 doublet that are also
formed at the wind base.

Closer to the stellar surface the rotation effects on H$\alpha$ diminish.
At this depth, the behavior of narrow lines becomes interesting.
The iron lines  around \ion{C}{IV}~$\lambda\lambda$1548--1552 doublet
are broadened and skewed to the blue.
This is the combined result of the large angular size of the stellar surface, limb
darkening, and the broken forward-backward symmetry in the azimuthal direction.
We will discuss this issue below in detail.
At stellar surface ($v_r \sim 0$) the optical depth is so large that any parcel of
material sees only its immediate neighborhood which roughly moves with the
same velocity.
Consequently, no skewness, displacement or line-shape difference occurs between
the profiles calculated for different latitudes (not shown in Figs.~\ref{fig9} and \ref{fig10}).

Figure~\ref{fig11} shows the detailed structure of the \ion{He}{I}~4713.17~\AA\ profile
in $J$ moment.
Since this line is not affected by blending  (see e.g., the second
panel of Fig.~\ref{fig12}), its position, shape, and width should clearly reflect the
expected rotation effects and should highlight any inconsistencies in our model
calculation.
We present these profiles in velocity space and correct for the local radial velocities.
The bottom row of  Fig.~\ref{fig11} shows \ion{He}{I}~4713.17~\AA\  deep in the
atmosphere ($v_r \sim$ 0 and $\tau_{\nu}$~$>>$~1).
The  line is in weak emission centered around 0~km~s$^{-1}$ as expected.
The profiles are similar at all latitudes which reflects the fact that only radiation from
the nearby co-moving regions contributes to $J$ at this position.
The line width reflects the local turbulent velocity and temperature.
The top row of Fig.~\ref{fig11} shows the normalized $J$ at $v_r \sim V_{\infty}$.
Here the line is in absorption and the profile widths show strong latitudinal dependence.
We expect \ion{He}{I}~4713.17~\AA\ to form in the photosphere, far from the
radii where $v_r \sim v_{\infty}$ ($r \sim 50 r_{core}$).
In the co-moving frame of this position the central star covers only a small solid angle
on the sky and can be considered as moving away with a uniform velocity, roughly equal
to $v_{\infty}$.
When we correct for the radial velocity of this position, we almost correctly account
for the Doppler shift of each small section of the photosphere, hence, the profiles in
Fig.~\ref{fig11} should be and are centered on $\sim 0$~km~s$^{-1}$.
The polar view (solid line) shows the intrinsic line profile (unaffected by rotation)
while the equatorial view (dash-dotted)  broadened by $\pm$250~km~s$^{-1}$ as
it should.

The profiles displayed in the middle panel of Fig.~\ref{fig11} are more difficult to
understand.
They appear to be blueshifted and also skewed at $\beta$= $\frac{\pi}{4}$ and
$\frac{\pi}{2}$.
At these intermediate radii ($v_r \sim 0.1v_{\infty}$ and $r \sim 1.5 r_{core}$) the
stellar surface covers a large portion of the sky and the Doppler shifts of photospheric
regions vary substantially.
The line profile in $J$ is a superposition of the profiles emanating from different
photospheric regions, and it is affected by the angular size of the photosphere and
by the limb darkening.
The line center should be redshifted by less than $0.1v_{\infty}$ velocity
which explains the $\sim$~$-$20~km~s$^{-1}$ blueshift in the middle panel of
Fig~\ref{fig11} (i.e., we over compensated the Doppler shift).
The blueward tilt of the profiles at $\beta= \frac{\pi}{4}$ and $\frac{\pi}{2}$ is
caused by the forward-backward asymmetry around the rotational axis.
The trailing and leading side of the photosphere contributes a broader and narrower
profile, respectively, which causes the blueward tilt.
We can conclude, therefore, that the gross characteristics of the
\ion{He}{I}~4713.17 \AA\  line profiles in Fig.~\ref{fig11} reflect the expected features
at all depths and reveal no inconsistencies in our method.

Figure~\ref{fig12} shows the observed spectra at different viewing angles around
selected transitions.
We also show the calculations of CMF\_FLUX for the corresponding spherical model.
Not surprisingly, the observed spectra reveal the same characteristics as those of
$J$ moment at large radii.
For our purposes, the most important feature of Figs.~\ref{fig12} is the remarkable
similarity to Figs.~4 and 5 of \cite{bus05}.
Despite the limited ability of our code to produce observed
spectra, Fig.~\ref{fig12} shows all the qualitative features of the synthetic
observations.
Most of the differences are due to our treatment of the ES.
Our code does not redistribute the scattered radiation in frequency space
which would produce smoother features like those of \cite{bus05}.
Note, that we run CMF\_FLUX with coherent ES for proper comparison;
therefore, the spherical symmetric spectra also show sharper features.

\section{Summary}\label{section:con}

We have implemented the short-characteristic method into a radiation
transfer code that can handle axi-symmetric stellar models with realistic
wind-flow velocities.
This routine will replace the continuum transfer plus Sobolev
approximation approach that is currently used in our axi-symmetric
stellar atmosphere program \citepalias[ASTAROTH,][]{geo05}.
The new transfer code allows for non-monotonic wind-flow and, therefore, will
enhance ASTAROTH's ability to treat line transfer accurately in models
for Be stars, OB rotators, binaries with colliding winds or accretion disks,
pre-main sequence and young stars, and for collapsing core (Type-II) supernovae.

The most important improvements of our approach are the sampling
method that we introduced to map the directional variation of the
radiation, and the flexible approach to allow for non-monotonic
velocity fields.
We use a global grid in impact-parameters and in inclination angles
(the angle between the equator and the plane containing the ray and the origin),
and solve the transfer independently for every pair of these parameters.
The code calculates the incoming intensities for the characteristics -- a necessary
feature of the short-characteristic method -- by a single latitudinal interpolation.
Our approach eliminates the need for further interpolations in the radiation angles.
The effects of the wind-flow are taken into account by adapting the resolution
along the characteristics to the gradient of the flow velocity.
This method ensures the proper frequency mapping of the opacities and emissivities where
it is needed, but avoids performing unnecessary work elsewhere.
Furthermore, it also provides flexibility in trading accuracy for speed.

The code also allows for distributed calculations.
The work-load can be shared between the processors by either distributing
the impact-parameter -- inclination angle pairs for which the transfer is calculated
or by assigning different frequency ranges to the processors.

We tested our code on static 1D/2D pure scattering problems.
In all cases, it reproduced the reference result with an error of a few
percent.
More complex tests on realistic stellar envelopes, with and without rotation,
were also performed.
Our code reproduced the results of a well-tested 1D code \citep[CMFGEN,][]{hil98},
as well as the expected features in 2D rotating atmospheres.
These tests demonstrated the feasibility and accuracy of our method.
In a subsequent paper, we will describe the implementation of
our code into ASTAROTH and present the results of fully
self-consistent 2D simulations.

\begin{acknowledgements}

This research was supported by NSF grant AST-9987390.

\end{acknowledgements}


\appendix

\section{Interpolation Methods}\label{section:appendixA}

\subsection{Linear Interpolations}

We used bi-linear interpolations to calculate opacity, source function, and line of
sight velocity at non-grid positions in our modeling domain.
The values were calculated by a weighted average of the corresponding quantities
at the nearest grid points by
\begin{equation}\label{eq:bichi}
\chi (\nu) =  \sum_{l=1}^4 w_l \cdot \chi_l (\nu) \; ,
\end{equation}
\begin{equation}\label{eq:biS}
S (\nu) = \sum_{l=1}^4 w_l \cdot S_l (\nu)  \; ,
\end{equation}
and
\begin{equation}\label{eq:biv}
\underline{\bf n} \cdot {\bf v}= \sum_{l=1}^4 w_l \cdot \left( \underline{\bf n} \cdot
{\bf v}_l  \right) \;.
\end{equation}
The nearest grid points are described as
$l$= 1 ($r_1$, $\beta_1$), $l$= 2 ($r_1$, $\beta_2$), $l$= 3 ($r_2$, $\beta_1$),
$l$= 4 ($r_2$, $\beta_2$); which set the weights in Eqs.~\ref{eq:bichi}--\ref{eq:biv}
to
\begin{eqnarray}\label{eq:w}
w_1 = \frac{r-r_2}{r_1-r_2} \cdot \frac{\beta-\beta_2}{\beta_1-\beta_2}  &  &
w_2 = \frac{r-r_2}{r_1-r_2} \cdot \frac{\beta-\beta_1}{\beta_2-\beta_1}  \\
w_3 = \frac{r-r_1}{r_2-r_1} \cdot \frac{\beta-\beta_2}{\beta_1-\beta_2}  &   &
w_4 = \frac{r-r_1}{r_2-r_1} \cdot \frac{\beta-\beta_1}{\beta_2-\beta_1} \; .
\end{eqnarray}
Coordinates $r$ ($r_2 \ge r  > r_1$) and $\beta$ ($\beta_2 \ge \beta > \beta_1$)
are the coordinates of the general (non-grid) position.
Note, that the frequency dependent quantities were interpolated in the co-moving
frame!

Since the integrals in Eqs.~\ref{eq:tau} and \ref{eq:I} are evaluated in the
co-moving frame of the down-stream end point of the characteristic,
the opacities and source functions calculated by Eqs.~\ref{eq:bichi} and
\ref{eq:biS} need to be properly Doppler corrected for the evaluation of
the integrals.
For a frequency $\nu$ in the co-moving frame of the down-stream end point
the procedure goes as follows: a; first we find the Doppler shifts $\Delta z_j$
by Eq.~\ref{eq:biv} for every integration point j on the characteristics
(see \S\ref{section:solution} for definitions) b; we
find co-moving frequencies $\nu_k$  and $\nu_{k-1}$ so that
\begin{equation}\label{eq:freqs}
\nu_k  \ge \nu \cdot \left( 1 - \Delta z_j \right) > \nu_{k-1}
\end{equation}
at all integration points c; we find the opacity and source function for $\nu_k$ and
$\nu_{k-1}$ by
Eqs.~\ref{eq:bichi} and \ref{eq:biS} d; we use linear interpolation in frequency space to get
these parameters at $\nu \cdot \left( 1 - \Delta z_j \right)$ for the integrations.

This seemingly cumbersome procedure is actually a straightforward book-keeping that
can be programmed very efficiently in the presence of monotonic velocity fields.
Note, that we do not mean global monotonicity but a velocity field that is
monotonic along the SC!
One can assure such a situation by properly creating the spatial grid.

\subsection{Interpolation of the Intensities}\label{section:appendixB}

Linear interpolations of the intensities at the upstream end point of the SC
does not provide acceptable accuracy.
This is because of the accumulation of errors from all previous intensity
interpolations.
Extensive testing of our method showed that the best result was achieved by
using monotonic cubic interpolations \citep[e.g.,][]{ste90}.
We use this method to interpolate the intensities in $\beta$ angle
for fixed $r$ and $\nu$.
The monotonic cubic approximation provides the necessary 3rd order accuracy,
yet avoids artificial variations (``ringing'') that can be amplified and propagated
on our grid.
Using monotonic interpolations actually dampens out such ``ringings'' and
stabilizes our method.

An unfortunate effect of requiring monotonicity, however, is that we need to save the
intensities for all frequencies and $\beta$ angles on the previously treated shell
(see \S\ref{section:tests} for description of our code).
Fortunately, this does not impede our efforts to accommodate multi-processor
calculations because all ($p$,$i$) pairs can still be treated independently.
Only, we require an additional memory area for $\sim N_{\beta} N_{\nu}$ real
number per ($p$, $i$) pair.

\end{document}